\documentclass[twocolumn,aps,floatfix,showpacs,superscriptaddress]{revtex4}
\usepackage{amsmath,amssymb,eucal,graphicx}
\begin{document}
\title{Phase Transition with Non-Thermodynamic States in Reversible
Polymerization}
\author{E.~Ben-Naim}
\affiliation{Theoretical Division and Center for Nonlinear
Studies, Los Alamos National Laboratory, Los Alamos, New Mexico
87545 USA}
\author{P.~L.~Krapivsky}
\affiliation{Department of Physics,
Boston University, Boston, Massachusetts 02215 USA}
\begin{abstract}
We investigate a reversible polymerization process in which individual
polymers aggregate and fragment at a rate proportional to their
molecular weight.  We find a nonequilibrium phase transition despite
the fact that the dynamics are perfectly reversible.  When the
strength of the fragmentation process exceeds a critical threshold,
the system reaches a thermodynamic steady state where the total number
of polymers is proportional to the system size.  The polymer length
distribution has a sharp exponential tail in this case.  When the
strength of the fragmentation process falls below the critical
threshold, the steady state becomes non-thermodynamic as the total
number of polymers grows sub-linearly with the system size. Moreover,
the length distribution has an algebraic tail and the characteristic
exponent varies continuously with the fragmentation rate.
\end{abstract}


\maketitle

\section{Introduction}

Equilibrium systems relax to a steady state described by the Gibbs
distribution. In contrast, nonequilibrium systems are specified by the
dynamics rather then by a Hamiltonian, and there is no general
framework for describing nonequilibrium steady states.  Furthermore,
unlike equilibrium phase transitions that are characterized by robust
universality classes \cite{jes}, nonequilibrium phase transitions are
highly sensitive to details of the underlying dynamics \cite{hh}.

In this paper, we investigate polymerization dynamics, and we report
that competition between aggregation and fragmentation results in a
remarkable non-equilibrium phase transition. Despite the fact that the
dynamics are perfectly reversible, there is a nonequilibrium phase
transition from a thermodynamic state where the number of polymers is
proportional to the system size into a non-thermodynamic state where
the number of polymers is not proportional to the system size.

Reversible polymerization is ubiquitous in polymer and atmospheric
chemistry \cite{pjf,rld,sp}, and has analogies in networks \cite{ab}
and computer science \cite{bb,dja,jlr}.  Reversible polymerization
includes two competing processes: (i) The aggregation process
$[i]+[j]\to [i+j]$, merger of two polymer chains of lengths $i$ and
$j$ into a larger polymer, occurs with the aggregation rate $K_{ij}$;
(ii) The fragmentation process $[i+j] \to [i]+[j]$, breakage of a
polymer into two smaller polymers, proceeds with rate $F_{ij}$. This
process is reversible because the aggregation process and the
fragmentation process perfectly mirror each other.

Reversible polymerization is described by the master equations
\cite{note}
\begin{eqnarray}
\label{master}
\frac{dc_k}{dt}&=& \frac{1}{2}\sum_{i+j=k} K_{ij}\,c_i\,c_j-
c_k\!\sum_{j\geq 1} K_{kj}\,c_j\nonumber\\
&+&\sum_{j\geq 1}F_{kj}\,c_{j+k} - \frac{1}{2}\,c_k\!\sum_{i+j=k} F_{ij}
\end{eqnarray}
where $c_k(t)$ is the density of polymer chains composed of $k$
monomers at time $t$. The first two terms describe changes due to
aggregation and the next two terms account for changes due to
fragmentation.  The aggregation and fragmentation rates are
(non-negative) symmetric matrices, $K_{ij}=K_{ji}$ and
$F_{ij}=F_{ji}$.

In the simplest case, the steady state distribution is found by
equating the aggregation flux with the fragmentation flux,
\begin{equation}
\label{balance}
K_{ij}\,c_i\,c_j =  F_{ij}\,c_{i+j}.
\end{equation}
This {\em detailed balance} condition specifies an equilibrium state
where the fluxes between any two microscopic states of the system
balance. Such an equilibrium steady state exists for example when both
the aggregation and the fragmentation rates are constant
\cite{bt}. Another equilibrium state was found in a model of strings
at very high-temperatures with the rates $K_{ij}=ij$ and $F_{ij}=i+j$
\cite{lt}.

The detailed balance equation \eqref{balance} admits a solution only
when the aggregation and fragmentation rates satisfy special
relations, as shown in Appendix \ref{constraints}. In general, the
steady state distribution is specified by the full master equations
\eqref{master} and moreover, the detailed balance relations
\eqref{balance} may very well be violated.  For example, in a
``chipping'' process where only end-monomers can detach from the
polymer, the matrix $F_{ij}$ is sparse: $F_{ij}=0$ when both $i,j\geq
2$. For constant aggregation rates, the chipping process exhibits a
nonequilibrium phase transition.  When the fragmentation rate falls
below a certain threshold, a giant macroscopic polymer emerges
\cite{vzl,kr,mkb,rm}.

We consider the aggregation and fragmentation rates 
\begin{equation}
\label{rates}
K_{ij}=ij, \qquad F_{ij}=\lambda.
\end{equation}
These rates, while intermediate between the linear chain model
\cite{bt} and the string model \cite{lt}, violate detailed balance 
(see Appendix \ref{constraints}). The product aggregation rate
accounts for the natural situation in which any two monomers may form
a chemical bond, thereby leading to merger of their respective
polymers.  This polymerization process has been widely studied in
polymer chemistry \cite{pjf1,whs,jdb,jls,ve} and in the context of
percolation \cite{sa,fl,zhe,aal}.  The constant fragmentation rate
reflects situations where all chemical bonds in the linear polymer are
equally likely to break, thereby leading to breakage into two smaller
polymers.  This de-polymerization process has also been studied
extensively \cite{zm}. Like the aggregation rate, the total
fragmentation rate is linear in the molecular weight,
$\sum_{i+j=k}F_{ij}=\lambda(k-1)$.

Starting with $N$ monomers, we study the nonequilibrium steady states
that emerge in the reversible polymerization process \eqref{rates}. We
find that the system generally reaches a steady state, and that a
nonequilibrium phase transition occurs at the critical fragmentation
rate $\lambda_c=1$.  The average total number of polymers, $N_{\rm
tot}$, grows algebraically with the system size $N$,
\begin{equation}
\label{ntot}
N_{\rm tot}\sim N^{\gamma} \qquad {\rm with} \qquad
\gamma=\frac{2\lambda}{2+\lambda},
\end{equation}
when fragmentation is weak, $\lambda<\lambda_c$. The total number of
polymers grows sub-linearly with the system size because
$\gamma<1$. Moreover, large polymers are likely as the polymer size
distribution has a broad algebraic tail.  The system develops this
non-thermodynamic state through a gelation transition.  We probe this
gelation using moments of the size distribution.

In contrast, the system reaches an ordinary steady state when the
fragmentation process is strong. The average total number of polymers
is proportional to the system size, $N_{\rm tot}=(1-\lambda^{-1})N$,
when $\lambda>\lambda_c$. Large polymers become rare since the polymer
size distribution has a sharp exponential tail.

Interestingly, even though the polymerization process is reversible
because the underlying aggregation \hbox{($[i]+[j]\to [i+j]$)} and
fragmentation \hbox{($[i+j] \to [i]+[j]$)} processes perfectly mirror
each other and none of the transition rates \eqref{rates} vanish, the
breakdown of detailed balance leads to a remarkable phase transition
involving a non-thermodynamic phase where the number of polymers is
not proportional to the system size and a thermodynamic phase where
the number of polymers is proportional to the system size.

The rest of this paper is organized as follows.  The thermodynamic
steady states that occur under strong fragmentation are examined in
the next section, while the non-thermodynamic steady states that
emerge when fragmentation is weak are analyzed in section
\ref{nonthermo}.  The gelation transition is probed using the moments
of the size distribution in section \ref{gelation}. Monte Carlo
simulation results validating the theoretical predictions for the
non-thermodynamic phase are detailed in section \ref{simulations}.  We
discuss the results and several open-ended questions in section
\ref{discussion}. Appendices \ref{constraints}--\ref{infinite} contain
several technical derivations.

\section{Thermodynamic Phase}
\label{thermo}

Our focus is the steady state behavior and in particular, the
stationary polymer size density $c_k$ that satisfies
\begin{equation}
\label{ck:eq}
\frac{1}{2}\!\sum_{i+j=k} ij\,c_i c_j -  k\,c_k 
=-\lambda\sum_{j>k}^\infty c_j+\frac{\lambda}{2}\,(k-1)c_k.
\end{equation}
This steady state equation is obtained by substituting the aggregation
and fragmentation rates \eqref{rates} into the stationary master
equation \eqref{master}. At the steady state, changes due to
aggregation, represented on the left hand side, balance changes due to
fragmentation, represented on the right hand side.  Since both
aggregation and fragmentation do not alter the total mass, the overall
mass density $\sum_k k c_k$ is a conserved quantity, as follows from
the rate equation \eqref{master}. We conveniently set the
normalization $\sum_k k c_k=1$ without loss of generality.

The total polymer density $M_0=\sum_k c_k$ is the most elementary
probe for the state of the system. At the steady state, this quantity 
satisfies 
\begin{equation}
\label{M0:eq}
\frac{1}{2}=\frac{\lambda}{2}\left(1-M_0\right),
\end{equation}
an equation obtained by summing \eqref{ck:eq} and by using the identity
$\sum_k\sum_{j\geq k} c_j =\sum_j c_j\sum_{k<j}1=\sum_j (j-1)c_j$ and
the normalization condition $\sum_k kc_k=1$. The total density is
non-zero 
\begin{equation}
\label{M0:sol}
M_0=1-\lambda^{-1},
\end{equation}
when the fragmentation rate is sufficiently strong, $\lambda>1$.  We
focus on this strong fragmentation regime in the rest of this section.

Let us assume that the system is large but finite with a total mass
equal $N$, a state that can be achieved by starting with $N$ monomers,
for example.  The expected total number of polymers, $N_{\rm
tot}=N\sum_k c_k=NM_0$ is proportional to the system size $N$, and
therefore, the system is in a thermodynamic state.

The polymer size density can be calculated by utilizing the recurrent
nature \cite{rec_nat} of Equation \eqref{ck:eq}. For instance, the
monomer and the dimer densities are
\begin{subequations}
\begin{align}
\label{c1}
c_1 &= \frac{\lambda-1}{\lambda+1},\\
\label{c2}
c_2 &=\frac{(\lambda-1)(3\lambda+1)}{(\lambda+1)^2(3\lambda+4)}.
\end{align}
\end{subequations}
In general, the polymer size density is finite in the thermodynamic
phase, $\lambda>1$. Very large polymers are rare since the size
distribution decays exponentially
\begin{equation}
\label{ck:asymp}
c_k \simeq Ak^{-5/2}\,e^{-ak}, 
\end{equation}
when $k\to\infty$.  This result is derived in Appendix \ref{derivation}.

When fragmentation is extremely strong, the system consists primarily
of monomers: $c_1= 1+O(\lambda^{-1})$ and $c_2=
\lambda^{-1}+O(\lambda^{-2})$ when $\lambda\to \infty$. The leading
asymptotic behavior can be obtained exactly in this strong
fragmentation limit,
\begin{equation}
\label{ck:sol}
c_k \simeq
\frac{k^{k-2}}{k!}\left(\frac{2}{\lambda}\right)^{k-1},
\end{equation}
for all $k$. This expression, obtained in Appendix \ref{infinite}, is
compatible with the generic exponential tail \eqref{ck:asymp}.

\subsection*{The near-critical behavior}

The total density, the monomer density, and the dimer density all
vanish near the transition point, $M_0\simeq (\lambda-1)$, $c_1\simeq
\frac{1}{2}(\lambda-1)$, and $c_2\simeq \frac{1}{7}(\lambda-1)$ as
$\lambda\to 1$.  This behavior suggests the perturbative approach, 
\begin{equation}
\label{ck:sub}
c_k = \epsilon \, b_k 
\end{equation}
with the small parameter $\epsilon = \lambda-1$. The first two
coefficients are $b_1=\frac{1}{2}$ and $b_2=\frac{1}{7}$.  We
substitute this form into the stationary equation \eqref{ck:eq} and
observe that the nonlinear aggregation term $\propto \epsilon^2$ is
negligible. Consequently, to leading order, the polymer size density
obeys the {\it linear} recursion equations
\begin{subequations}
\begin{align}
k\,b_k&=\sum_{j=k+1}^\infty b_j-\frac{1}{2}\,(k-1)b_k,\\
(k+1)b_{k+1}&=\sum_{j=k+2}^\infty b_j-\frac{1}{2}\,k\,b_{k+1}.
\end{align}
\end{subequations}
The second equation is obtained from the first by an index shift.  We
subtract the two equations and obtain a recursion relation for the
coefficients $b_k$,
\begin{equation}
\label{bk:rec}
\frac{b_{k+1}}{b_k} = \frac{k- \frac{1}{3}}{k+\frac{4}{3}}.
\end{equation}
The coefficients can be conveniently expressed as a ratio of Gamma
functions, $b_k\propto \Gamma(k-1/3)/\Gamma(k+4/3)$, by using the
identity $\Gamma(x+1)/\Gamma(x)=x$. The polymer size density is
therefore
\begin{equation}
\label{bk:sol}
b_k = \frac{1}{2}
\frac{\Gamma(\frac{7}{3})}{\Gamma(\frac{2}{3})}
\frac{\Gamma(k-\frac{1}{3})}{\Gamma(k+\frac{4}{3})},
\end{equation}
where the proportionality constant is set by \hbox{$b_1=\frac{1}{2}$}.

Near criticality, the size density is algebraic,
\begin{equation}
\label{ck:small}
c_k\sim \epsilon \, k^{-5/3},
\end{equation}
over a substantial range, $k\ll k_*$. This result follows from
\eqref{bk:sol} and $\lim_{x\to\infty} x^{a}\Gamma(x)/\Gamma(x+a)= 1$.
Therefore, the likelihood of finding large polymers becomes
substantial as the phase transition point is approached.  The cutoff
scale $k_*$, set by mass conservation, $\sum_{k=1}^{k_*}kc_k=1$, is
divergent
\begin{equation}
\label{cutoff}
k_*\sim \epsilon^{-3}\,.
\end{equation}
The size distribution is sharply suppressed according to
\eqref{ck:asymp} beyond this scale. Using the relation $A\sim
a^{-1/2}$, derived in appendix \ref{derivation}, and $a\sim k_{*}^{-1}$
we deduce that
\begin{equation}
\label{ck:large}
c_k\sim \epsilon^{-3/2} k^{-5/2}e^{-{\rm const}\times{\epsilon^3} \,k}
\end{equation}
for $k\gg k_*$. Indeed, this large size behavior matches the small
size behavior \eqref{ck:small} at the crossover scale \eqref{cutoff}.
We conclude that the convolution term, that accounts for the creation
of very large polymers from smaller polymers, is relevant only at very
large scales, $k\gg k_*$. Otherwise, this term does not affect the
density of small polymers.
  
For completeness, we mention that the leading asymptotic behavior of
the moments, $M_n=\sum_k k^n c_k$, readily follows from the density
\eqref{ck:small} and the cutoff \eqref{cutoff},
\begin{equation}
\label{Mn:nc}
M_n\sim
\begin{cases}
\epsilon^{-3(n-1)}&n>2/3\\
1&n<2/3.
\end{cases}
\end{equation}
Sufficiently large order moments diverge in the vicinity of the
transition point, a consequence of the algebraic tail
\eqref{ck:small}.  The low order moments are finite, however.

\section{Non-thermodynamic Phase}
\label{nonthermo}

As the critical point is approached, the nonlinear convolution term in
\eqref{ck:eq} becomes irrelevant over the divergent scale
\eqref{cutoff}. By continuity, we deduce that the convolution term is
negligible when $\lambda<1$. Consequently, the stationary distribution
obeys the linear equation
\begin{equation}
\label{ck:lin}
k\,c_k =\lambda\left(\sum_{k=1}^\infty c_k-\sum_{j=1}^{k} c_j\right)
-\frac{\lambda}{2}\,(k-1)c_k
\end{equation}
when $\lambda<1$.  We introduce the normalized size density,
$\rho_k=c_k/\sum_k c_k$, with $\sum_{k\geq 1} \rho_k =1$. With this
transformation, the stationary equation \eqref{ck:lin} becomes
\begin{equation}
\label{rhok:lin}
k\,\rho_k =\lambda\left(1-\sum_{j=1}^k
\rho_j\right)-\frac{\lambda}{2}\,(k-1)\rho_k.
\end{equation}

The monomer and dimer densities follows immediately,
\begin{subequations}
\begin{align}
\label{rho1}
\rho_1&=\frac{\lambda}{1+\lambda},\\
\label{rho2}
\rho_2&=\frac{2\lambda}{(1+\lambda)(4+3\lambda)}.
\end{align}
\end{subequations}
The normalized densities undergo a phase transition at $\lambda_c=1$,
as shown in figure 1. The fraction of monomers is not affected by the
convolution term and \eqref{rho1} holds for all $\lambda$. However,
the dimer density \eqref{rho2} differs from the expression
\hbox{$\rho_2=\frac{\lambda(1+3\lambda)}{(1+\lambda)^2(4+3\lambda)}$}
for $\lambda<1$ implied by \eqref{c2} and \eqref{M0:sol}. Similarly,
the normalized size densities $\rho_k$ exhibit a phase transition for
all $k>1$. 

\begin{figure}[t]
\includegraphics[width=0.4\textwidth]{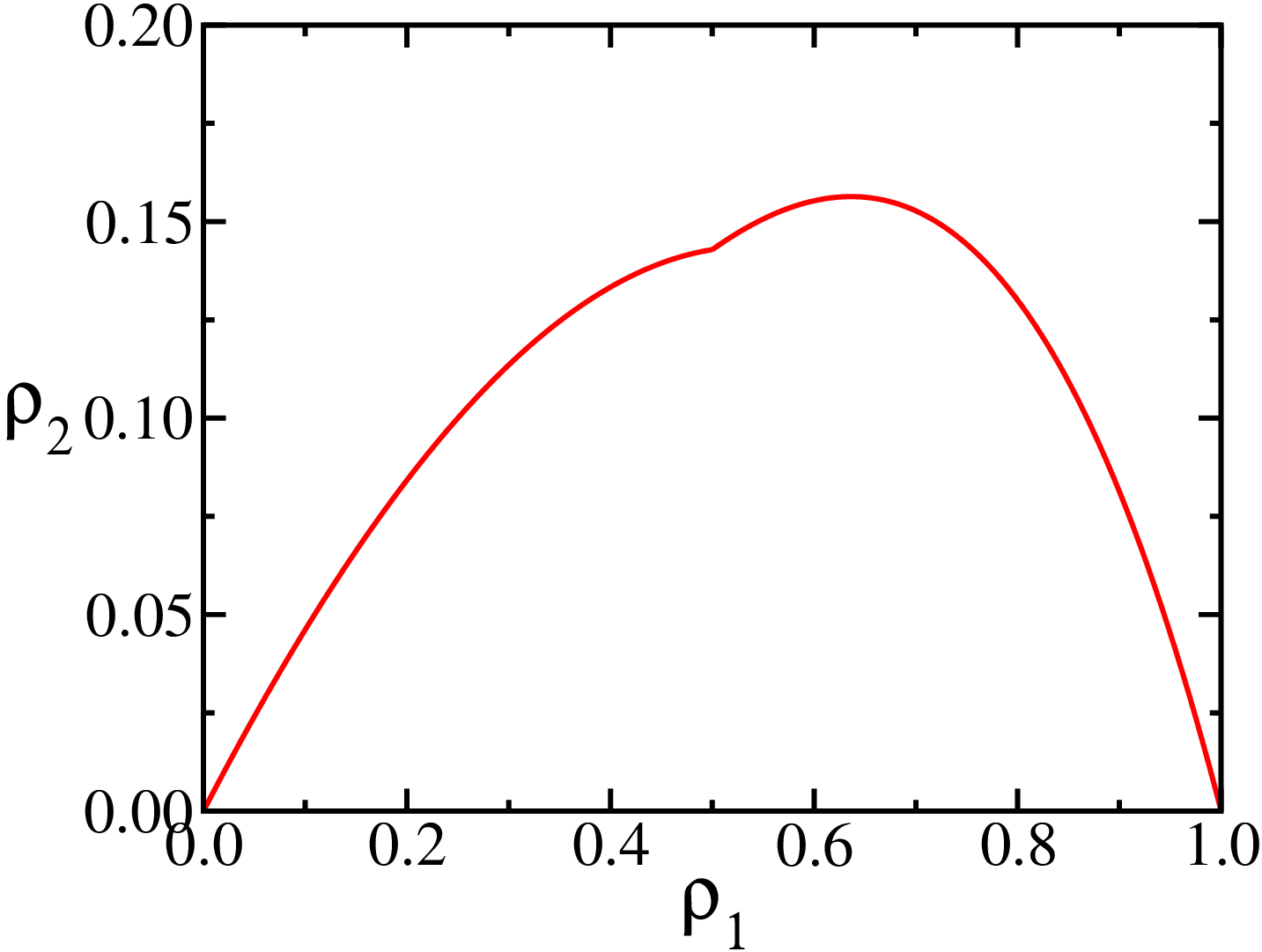}
\caption{The normalized dimer density $\rho_2$ versus the normalized
monomer density $\rho_1$. The phase transition at $\lambda_c=1$ is
reflected by the discontinuity in the first derivative at
$\rho_1=1/2$.}
\label{fig-rho}
\end{figure}

Generally, we recast Eq.~\eqref{rhok:lin} into the following recursion
for the normalized densities
\begin{equation}
\label{rhok:rec}
\frac{\rho_{k+1}}{\rho_k} = 
\frac{k- \frac{\lambda}{2+\lambda}}{k+\frac{2(1+\lambda)}{2+\lambda}}.
\end{equation}
This recursion is obtained by repeating the steps leading to
\eqref{bk:rec}. Again, we express the normalized densities as a ratio
of Gamma functions, \hbox{$\rho_k \propto
\Gamma(k-\frac{\lambda}{2+\lambda})/
\Gamma(k+\frac{2(1+\lambda)}{2+\lambda})$}. The monomer density
\eqref{rho1} sets the proportionality constant and hence,
\begin{equation}
\label{rhok:sol}
\rho_k= \frac{\lambda}{1+\lambda}
\frac{\Gamma\left(1+\frac{2(1+\lambda)}{2+\lambda}\right)}
{\Gamma\left(1- \frac{\lambda}{2+\lambda}\right)}
\frac{\Gamma\left(k- \frac{\lambda}{2+\lambda}\right)}
{\Gamma\left(k+\frac{2(1+\lambda)}{2+\lambda}\right)}.
\end{equation}
The size density has an algebraic tail,
\begin{equation}
\label{rhok:pl}
\rho_k\sim k^{-\beta} \qquad {\rm with} 
\qquad \beta=\frac{2+3\lambda}{2+\lambda},
\end{equation}
as $k\to\infty$, thereby implying that large polymers are likely. The
decay exponent $1<\beta<5/3$ is not universal.

The size density obeys $c_k\propto \rho_k$, and the $N$-dependent
proportionality constant is obtained from the mass conservation
condition $\sum_{k=1}^N k c_k=1$ where the upper limit of integration
is set by the system size. This sum is dominated by the density of
large polymers. By performing the summation, we find that the polymer
size density depends on the system size
\begin{equation}
\label{ck:pl}
c_k\sim N^{\beta-2} k^{-\beta}. 
\end{equation}
The total number of clusters $N_{\rm tot}\sim N\sum_k c_k$ grows
sub-linearly with the system size $N_{\rm tot}\sim N^{\gamma}$ with
$\gamma=\beta-1$ as announced in \eqref{ntot}. Therefore, the total
polymer density, $M_0\sim N^{\beta-2}$, depends on the system size in
contrast with the behavior when $\lambda>1$. In deriving the
steady state equation \eqref{ck:lin}, we assumed that the convolution
term is negligible. This assumption is consistent with the fact that
the amplitude $N^{\beta-2}$ in \eqref{ck:pl} vanishes as the system
size diverges. 

Moreover, the expected total number of polymer of size $k$,
$C_k=Nc_k$, is as follows, $C_k\sim N^{\gamma} k^{-\beta}$ with
$\gamma=\frac{2\lambda}{2+\lambda}$.  This steady state is not
thermodynamic!  The number of polymers is much smaller then the system
size, yet the number of polymers still diverges with the total mass: $C_k$
grows sub-linearly with the system size because $\gamma<1$.

For irreversible polymerization, $\lambda=0$, all mass eventually ends
up in a single giant polymer, as reflected by the characteristic
exponent $\gamma=0$.  The total number of polymers still grows
sub-linearly, $N_{\rm tot}\sim N^{2/3}$, when the critical point is
approached, $\lambda\to 1$.

In the thermodynamic phase, all polymers are finite in size. Indeed,
the exponential tail behavior \eqref{ck:asymp} implies that the
largest polymers are finite in scale.  Near criticality, the scale of
the largest polymers diverges according to \eqref{cutoff}. In the
non-thermodynamic phase, there are polymers of all possible scales
because the power-law behavior \eqref{ck:pl} holds up to the system
size $\propto N$.  Remarkably, there are polymers that contain a
finite fraction of all the mass in the system because according to
\eqref{ck:pl}, the total number of macroscopic clusters, $N\sum_{k\geq
{\rm const.}\times N} c_k $ is of the order one.

The power-law distribution \eqref{rhok:pl} accounts for a competition
between two fluxes. There is a flux of mass from small scales to large
scales that is generated by the aggregation process and a flux from
large scale to small scales caused by fragmentation. The power-law
behavior holds for all scales, indicating that these two fluxes
balance at all intermediate scales. Similar competitions between the
fluxes occur in fluid turbulence \cite{fluid}, passive scalar
advection \cite{passive}, wave turbulence \cite{zlf}, granular gases
\cite{bm}, and driven aggregation systems \cite{ht,kmr,crz}. However,
reversible polymerization differs from these driven system in that
there is no external injection of mass to maintain the steady-state.

\section{The Gelation Transition}
\label{gelation}

We now study the approach toward the steady state specified by
the full master equation
\begin{eqnarray}
\label{ckt:eq}
\frac{dc_k}{dt} = \frac{1}{2}\!\!\sum_{i+j=k} \!\!ij\,c_i c_j - k\,c_k +
\lambda\!\sum_{j>k}^\infty \!c_j -\frac{\lambda}{2}\,(k-1)c_k.
\end{eqnarray}
Initially, there are only monomers, \hbox{$c_k(t=0) = \delta_{k,1}$}.

First, we consider the total polymer density $M_0$ that obeys the
linear rate equation
\begin{eqnarray}
\label{M0t:eq}
\frac{dM_0}{dt} = -\frac{1}{2} + \frac{\lambda}{2}\,(1-M_0).
\end{eqnarray}
Subject to the initial condition $M_0(0)=1$, the total density is 
\begin{equation}
\label{M0t:sol}
M_0 = 1 - \lambda^{-1}+\lambda^{-1}\,e^{-\lambda t/2}.
\end{equation}
Hence, the steady state \eqref{M0:sol} is approached exponentially
fast in the thermodynamic phase. Moreover, the monomer and the dimer
densities also relax exponentially fast as in \eqref{M0t:sol} and in
general, the polymer size density quickly approaches the steady state
when $\lambda>1$.

We focus on the kinetics in the more interesting non-thermodynamic
phase where the total polymer density \eqref{M0t:sol} vanishes at 
time $t_0=\frac{2}{\lambda}\ln \frac{1}{1-\lambda}$.  This
behavior is consistent with the vanishing total density, $M_0\sim
N^{\beta-2}$, implied by \eqref{ck:pl}.  Of course, the negative
expression \eqref{M0t:sol} is invalid beyond the time $t_0$.

The moments $M_n=\sum_k k^nc_k$ provide a direct probe of the
kinetics. In the non-thermodynamic phase, the system nucleates large
macroscopic gels and consequently, large moments diverge with the
system size as follows from \eqref{ck:pl}. This, together with the
vanishing overall density $M_0$, indicates that the system undergoes a
gelation transition at a finite time.  At the gelation time $t_g$, a
giant polymer or a gel emerges as is the case for irreversible
polymerization ($\lambda=0$).  From the master equation
\eqref{ckt:eq}, the moments evolve according to
\begin{eqnarray}
\label{AF:Mn}
\frac{dM_n}{dt} &=& \frac{1}{2}\sum_{m=1}^{n-1}\binom{n}{m}M_{m+1}M_{n+1-m}
- \frac{\lambda}{2}\,\frac{n-1}{n+1}\,M_{n+1}\nonumber\\
&+&\frac{\lambda}{n+1}
\sum_{m=2}^{n}\binom{n+1}{m}B_{m}M_{n+1-m}
\end{eqnarray}
where $B_m$ are the Bernoulli numbers \cite{gkp}. For example, the
second moment obeys $dM_n/dt=M_2^2-\frac{\lambda}{6}(M_3-1)$.  We
assume that large order moments diverge algebraically at the gelation
time, \hbox{$M_n\sim(t_g-t)^{-(an+b)}$} for $n\geq 1$, and observe
that the last term in the hierarchical equation \eqref{AF:Mn} is
negligible compared with the rest of the terms. We require that the
time dependent term, the aggregation term, and the remaining
fragmentation term are comparable and find that $a=-b=1$. Hence,
\begin{equation}
\label{Mn:diverge}
M_n\sim (t_g-t)^{-(n-1)}
\end{equation}
for $n\geq 1$.  Indeed, the moments diverge at a finite time. The
exponent $n-1$ characterizing this divergence is compatible with the
near critical behavior \eqref{Mn:nc}.  The divergence
\eqref{Mn:diverge} is different than the $M_n\sim (t_g-t)^{-(2n-3)}$
behavior in irreversible fragmentation \cite{fl}, and therefore,
fragmentation quantitatively alters the nature of the gelation
transition.

We conclude that the solution to the master equation \eqref{ckt:eq}
exhibits a finite time singularity. Even though we can not obtain the
gelation time $t_g$ exactly, relevant properties of the size density
including the moments can still be obtained analytically. In
particular, the form of the size density at the gelation time can be
calculated by balancing the fluxes of mass due to aggregation and
fragmentation, following the scaling analysis in ref.~\cite{ve1}.

Consider $M^{(n)}$, the total mass density of polymers with size
smaller then $n$,
\begin{equation}
\label{mn:def}
M^{(n)}(t) = \sum_{k=1}^n kc_k(t). 
\end{equation}
According to the master equation \eqref{ckt:eq}, this quantity obeys
\begin{equation}
\label{mn:flux}
\frac{dM^{(n)}}{dt} \!= \!
-\sum_{i=1}^n \!\sum_{j+i=n-1}^\infty \!\!i^2j c_ic_j
+\frac{1}{2}\,\lambda n(n+1)\!\!\!\sum_{j=n+1}^\infty c_j.
\end{equation}
We now take the $n\to\infty$ limit. The aggregation loss term accounts
for loss of finite size polymers to the infinitely large gel while the
fragmentation term accounts for the balancing flux from the gel into
small masses.  We require that the two fluxes balance at the gelation
point.  We assume that at the gelation point, the size density decays
algebraically
\begin{equation}
\label{tau}
c_k(t=t_g) \sim k^{-\tau},
\end{equation}
for $k\gg 1$, as is the case for irreversible polymerization
\cite{zhe}. By dimensional counting, the aggregation flux term scales
as $n^{5-2\tau}$ while the fragmentation flux scales as
$n^{3-\tau}$. The two fluxes balance when $5-2\tau=3-\tau$ and as a
result $\tau=2$.  Therefore, $c_k(t=t_g)\sim k^{-2}$, a behavior that
is consistent with the aforementioned divergence of the moments
\eqref{Mn:diverge} when the power-law behavior \eqref{tau} holds up to
a cutoff scale that diverges near gelation, $k\ll (t_g-t)^{-1}$.

The normalization condition $\sum_k kc_k=1$ imposes the exponent
restriction $\tau>2$ but the heuristic argument above yields precisely
the marginal value $\tau=2$. We therefore anticipate that there is a
logarithmic correction with the following form $c_k\sim k^{-2}(\ln
k)^{-\mu}$. We substitute this form into \eqref{mn:flux} and then, the
aggregation term is of the order $n\,[\ln n]^{1-2\mu}$ while the
fragmentation term is of the order $n\,[\ln n]^{-\mu}$. Therefore, 
$\mu=1$, and 
\cite{nested}
\begin{equation}
\label{ck:tg}
c_k\sim k^{-2}[\ln k]^{-1}
\end{equation}
for $k\gg 1$.  This decay is milder then the \hbox{$c_k\sim k^{-5/2}$}
behavior found for irreversible polymerization \cite{zhe}, and
therefore, there are many more large clusters in reversible
polymerization. 

The size of the largest gel at the gelation transition follows
immediately from the extreme statistics criterion, $N\sum_{k=1}^{k_g}
c_k=1$. Remarkably, this size is nearly macroscopic in the size of the
system,
\begin{equation}
\label{kg}
k_g\sim N[\ln N]^{-1}.
\end{equation}
This gel size is much larger compared with the $k_g\sim N^{2/3}$
behavior in reversible polymerization. This increased scale enables
the gel to withstand fragmentation. We also comment that the nearly
macroscopic size scale \eqref{kg} provides the appropriate cutoff in
\eqref{mn:def}, $n\sim k_g$, and that at the gelation point, the
upward mass aggregation flux and the downward mass fragmentation flux
are nearly macroscopic, that is, they are proportional to the system
size up to a logarithmic correction.

\subsection*{The critical case}

For completeness, we discuss kinetics in the critical case $\lambda=1$
where the the cluster density \eqref{M0t:sol} is purely exponential,
$M_0 = e^{-t/2}$.  Similar decay characterizes the leading behavior of
the size density. For example, the monomer density obeys
\hbox{$dc_1/dt = -c_1+(M_0 -c_1)$} and consequently,
\begin{equation}
\label{c1:marginal}
c_1 = \frac{2}{3}\,e^{-t/2} + \frac{1}{3}\,e^{-2t}.
\end{equation}
Only the first term is relevant asymptotically, \hbox{$c_1\simeq
\frac{2}{3}\,e^{-t/2}$}. In general, $c_k\simeq u_k\,e^{-t/2}$, and by
substituting this expression into the time dependent master equation
\eqref{ckt:eq}, we observe that the nonlinear term is negligible.
Consequently, the coefficients $u_k$ satisfy the recursion equation
\begin{equation}
\left(k-\frac{1}{2}\right)u_k=\sum_{j=k+1}^\infty u_j -\frac{1}{2}(k-1)u_k. 
\end{equation}
{}From this recursion, the coefficients $u_k$ satisfy 
$u_{k+1}/u_k=(k-2/3)/(k+1)$. Therefore, the leading behavior of the
size density is as follows
\begin{equation}
u_k=\frac{2}{3\Gamma(\frac{1}{3})}
\frac{\Gamma(k-\frac{2}{3})}{\Gamma(k+1)}.
\end{equation}
The tail of the size density matches the near critical behavior
\eqref{ck:small}, $c_k\sim e^{-t/2}k^{-5/3}$. 

\section{Numerical Simulations}
\label{simulations}

We performed numerical simulations to validate the theoretical
predictions. Below, we present results for the non-thermodynamic
phase.

\begin{figure}[b]
\includegraphics[width=0.4\textwidth]{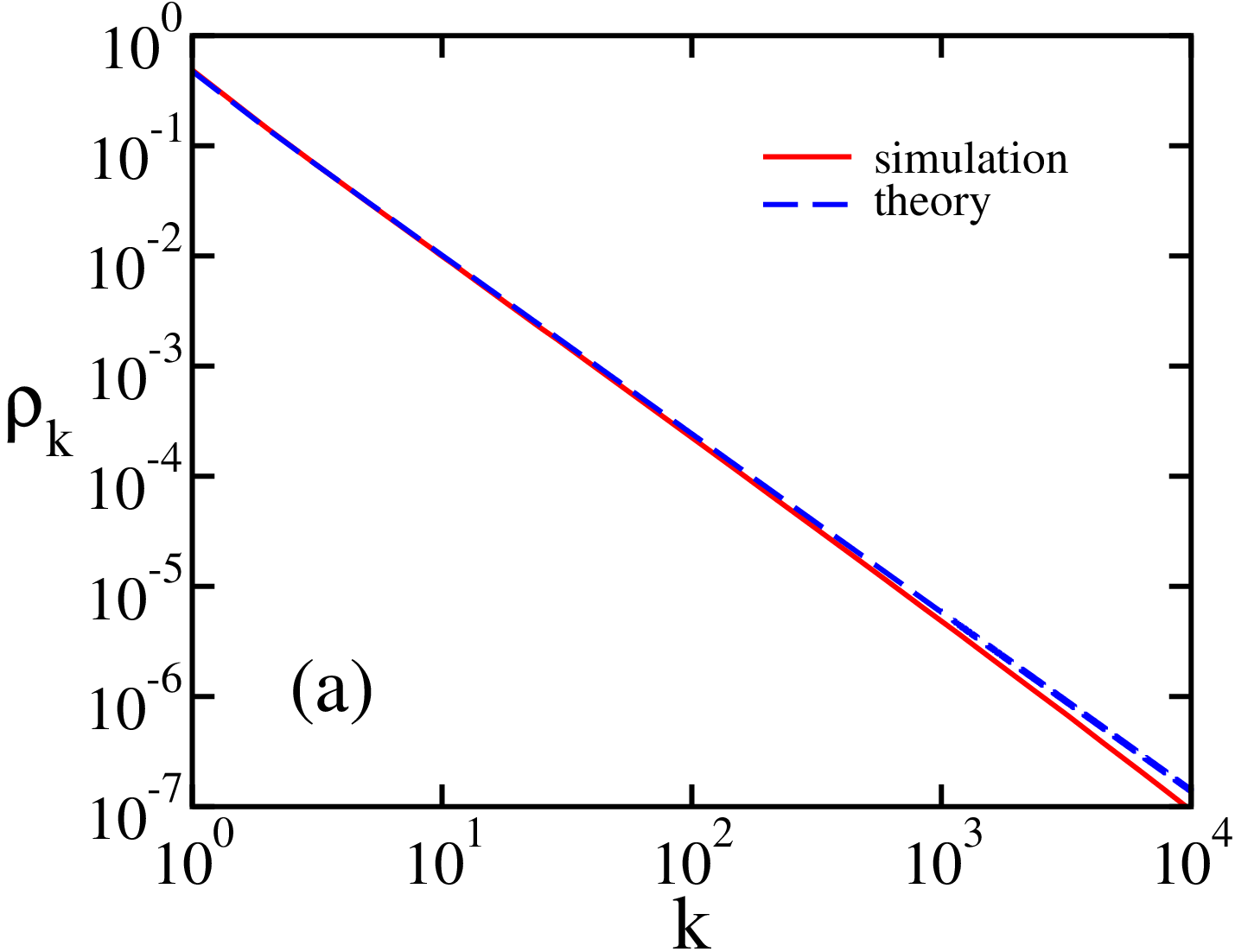}\vspace{.275in}
\includegraphics[width=0.4\textwidth]{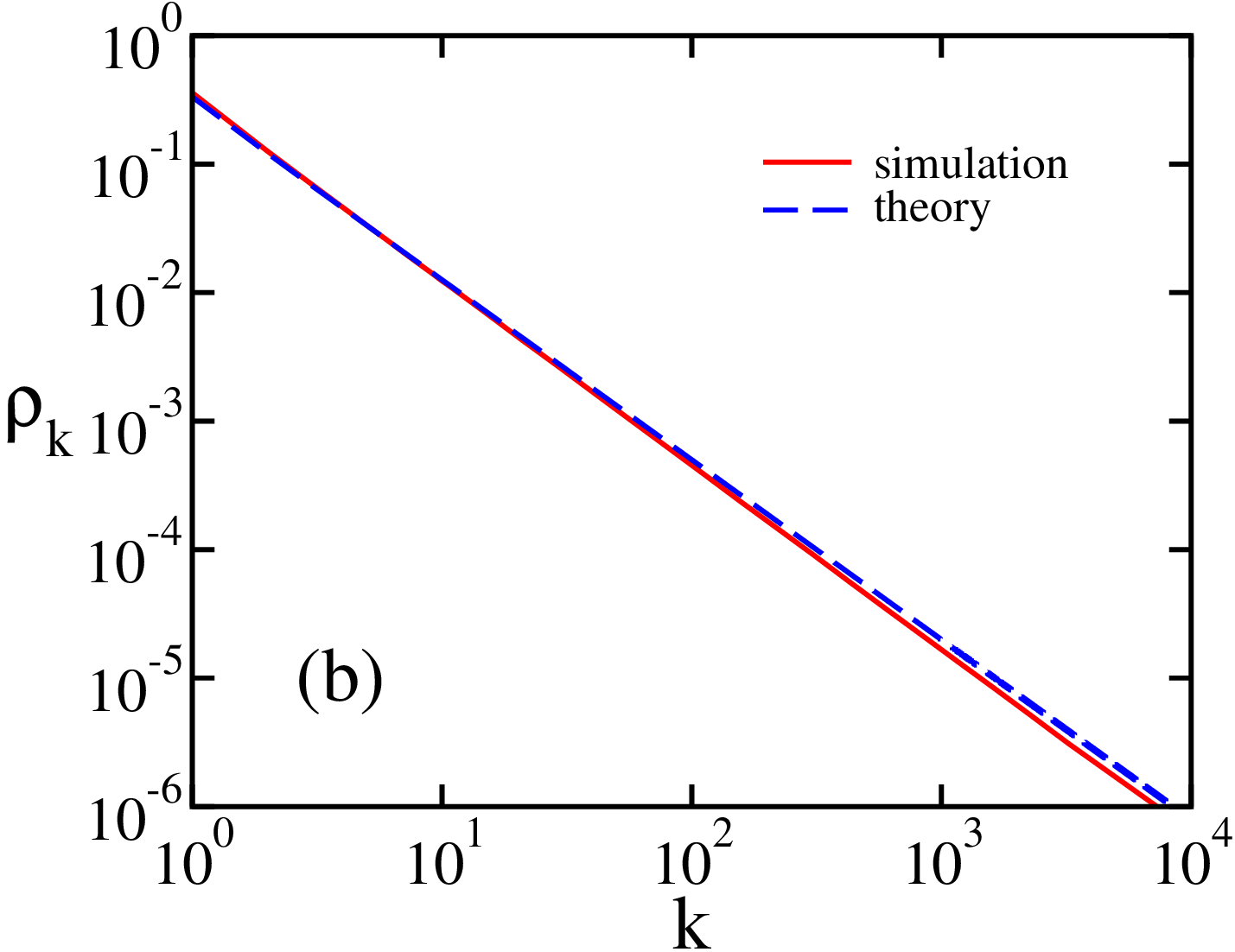}
\caption{The size distribution $\rho_k$ versus $k$ for $\lambda=0.9$
(a) and $\lambda=0.5$ (b). The simulations results are for a
system of size $N=10^5$.}
\label{fig-ck}
\end{figure}

The simulations were performed by starting with $N$ monomers and were
carried by repeating the following Monte Carlo step.  At each step,
the total aggregation rate $R_a=\frac{1}{N}\sum_{i\neq j} s_i s_j$ and
the total fragmentation rate $R_f=2\sum_i (s_i-1)$ are calculated
where $s_i$ is the size of the $i$th polymer. Of course, both of these
rates are proportional to the system size, \hbox{$R_a\propto
R_f\propto N$}.  An aggregation event is executed with probability
\hbox{$R_a/(R_a+R_f)$}, while a fragmentation event is executed with
the complementary probability $R_f/(R_a+R_f)$. In each aggregation
event, one polymer is chosen with probability proportional to its size
and is merged with another polymer, also chosen with probability
proportional to its size. In a fragmentation event, a polymer,
randomly chosen with probability proportional to the number of its
bonds, is randomly split into two smaller polymers. Time is updated by
the inverse of the total rate $t\to t+(R_a+R_f)^{-1}$ after each Monte
Carlo step. As a check, we successfully reproduced the total polymer
density \eqref{M0:sol}.

In the non-thermodynamic phase, the system undergoes a gelation
transition and then relaxes to the steady state.  We run the
simulations until the system relaxed to the steady state and then
obtained the size distribution from a long series of measurements to
reduce statistical fluctuations. The simulations results are for
systems of size $N=10^5$. We present results for the normalized
densities $\rho_k$ predicted in \eqref{rhok:sol}.  Overall, there is
very good agreement between the theoretical predictions and the
simulation results as shown in figure 2. The size distribution agrees
with \eqref{rhok:sol} and the tail of the distribution follows a
power-law as in \eqref{rhok:pl}. The simulation results agree with the
theoretical results slightly better near the phase transition point
(Figures 2a and 2b). Since the total number of clusters grows
sub-linearly with the system size, $N_{\rm tot}\sim N^\gamma$ with
$\gamma<1$, extremely large systems are needed to reduce the magnitude
of the statistical fluctuations. Such fluctuations are most pronounced
at the tail region where the discrepancy between the theory and the
simulation is a result of the limited system size.

\begin{figure}[t]
\includegraphics[width=0.38\textwidth]{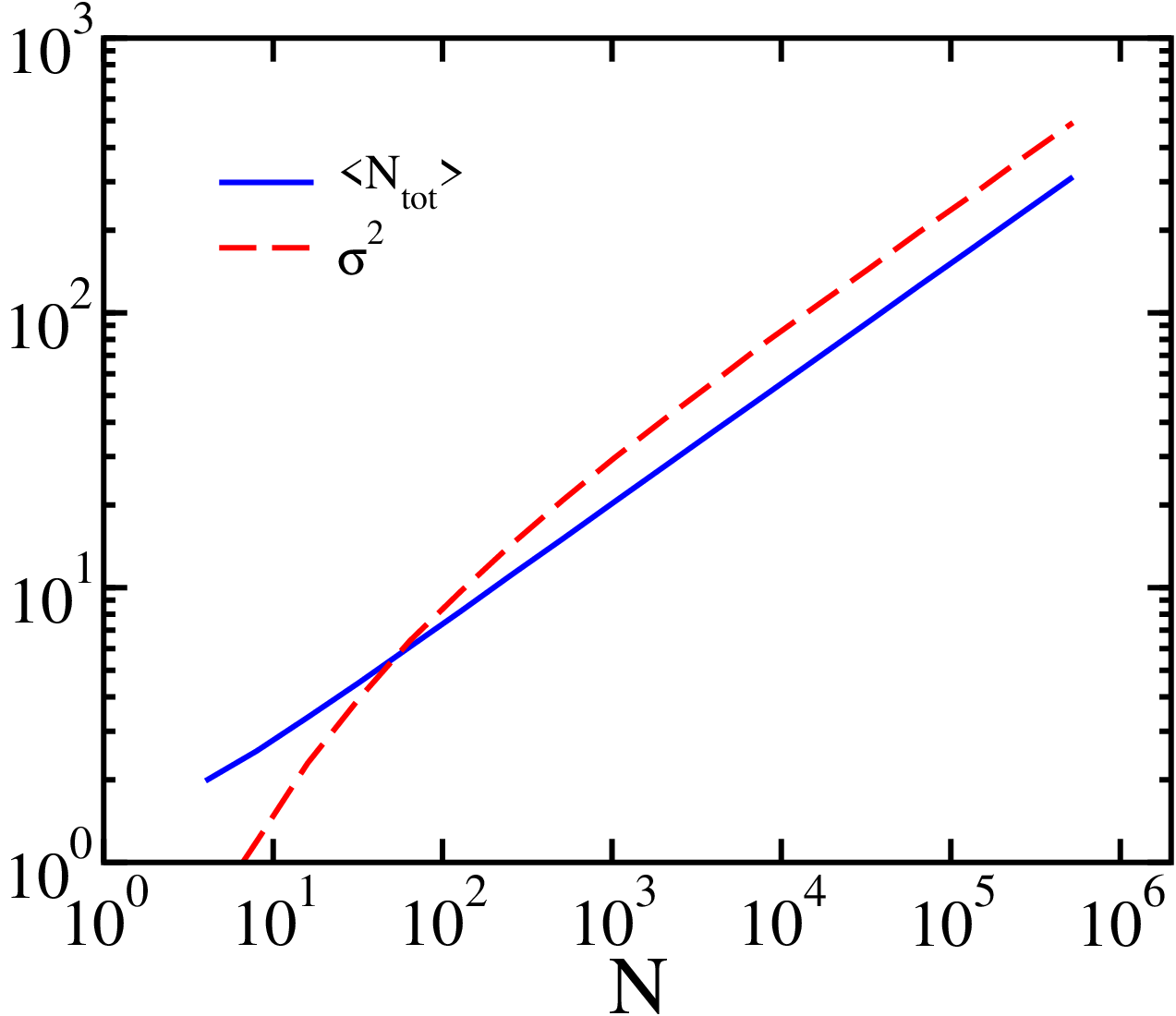}
\caption{Fluctuations in the total number of clusters. Shown are the
average number of clusters, $\langle N_{\rm tot}\rangle$ and the
variance, $\sigma^2=\langle N_{\rm tot}^2\rangle -\langle N_{\rm
tot}\rangle^2$ for $\lambda=0.5$.}
\label{fig-fluc}
\end{figure}

To quantify fluctuations in the total number of polymers, we also
measured the variance $\sigma^2=\langle N_{\rm tot}^2\rangle-\langle
N_{\rm tot}\rangle^2$. We find that the fluctuations follow a
central-limit like behavior \hbox{$\sigma^2\sim \langle N_{\rm
tot}\rangle$} (figure 3) and therefore,
\begin{equation}
\label{sigma}
\sigma^2\sim N^\gamma \qquad {\rm with} \qquad
\gamma=\frac{2\lambda}{\lambda+2}
\end{equation}
as in \eqref{ntot}.

\section{Discussion}
\label{discussion}

In summary, we studied stationary and dynamical properties of
reversible polymerization. We found an interesting phase transition
involving a thermodynamic phase and a non-thermodynamic phase. When
fragmentation is strong, the system is in a thermodynamic phase and
the number of polymers is proportional to the system size. The system
includes a large number of small polymers. When fragmentation is weak,
the system is in a non-thermodynamic phase as the total number of
polymers grows sub-linearly with the system size. In this phase, there
is a small number of large polymers since the size distribution is
power-law. Moreover, the polymer sizes are distributed at all
scales. Macroscopic gels may exist as well.

In the thermodynamic phase, the system quickly approaches the steady
state.  The time-dependent behavior is much richer in the
non-thermodynamic phase. The system exhibits a finite-time
singularity: large moments of the size distribution diverge at the
finite gelation time. At this time, the system nucleates macroscopic
gels and the size distribution follows a universal algebraic
decay. Past the gelation time, there is a second relaxation stage
leading the system to a state where there are two balancing fluxes of
mass: aggregation transfers mass from small scales to large scales and
fragmentation transfers mass from large scales to small scales. The
stationary size distribution is algebraic but the characteristic
exponent is not universal.

Even though the aggregation process is described by non-linear terms,
the analysis in the non-thermodynamic phase involves linear equations
because the aggregation gain term is relevant only at the largest
scale. Nevertheless, formation of gels at the largest possible scale
(in our case, the system size) is crucial in maintaining a stationary
state. Understanding the distribution of macroscopic gels is an open
challenge, and our numerical simulations reveal an interesting anomaly
with an enhancement of the population of macroscopic gels over the
algebraic distribution \eqref{ck:pl} at the maximal scale.

Missing from our calculations is a finite-size scaling analysis
\cite{bbckw,bk,ks,bk1} near the phase transition point. Interestingly,
the total number of clusters grows sub-linearly, \hbox{$N_{\rm
tot}\sim C(\lambda)\, N^{2/3}$}, just below the phase transition
point, but linearly just above the phase transition point,
\hbox{$N_{\rm tot}\sim (\lambda-1)N$}. Therefore, the amplitude
$C(\lambda)$ may very well be divergent, and moreover, there must be
an intermediate range of fragmentation rates centered around the
critical point with a smooth crossover between the two phases. The
width of this transition region should vanish as the system size
increases.

Understanding fluctuations is another interesting direction for
further research. We find Gaussian fluctuations in the thermodynamic
phase and are able to compute the variance
$\sigma^2=\frac{2}{\lambda}N$ but we are unable to obtain
\eqref{sigma} in the non-thermodynamic phase.

Finally, we checked that the same phase transition generally holds as long as
the aggregation rate is asymptotically proportional to the molecular
weights; for instance, when $K_{ij} = Aij + B(i+j) + C$, a class of
models that often arise in polymer chemistry \cite{pjf,rld}. A more
complete characterization of phase transitions in reversible
polymerization is another avenue for future work.

\acknowledgements
We acknowledge financial support from DOE grant DE-AC52-06NA25396 and
NSF grant CHE-0532969.

\appendix

\section{Detailed Balance}
\label{constraints}

In this appendix, we demonstrate that the detailed balance equations
\eqref{balance} do not generally have a solution.  The size densities 
$c_k$ satisfy 
\begin{eqnarray}
\label{c1234}
\begin{split}
K_{11} c_1^2    &= F_{11}c_2   \\
K_{12} c_1 c_2 &= F_{12}c_3   \\
K_{13} c_1 c_3 &= F_{13}c_4   \\
K_{22} c_2^2    &= F_{22}c_4   
\end{split}
\end{eqnarray}
for $k=1,2,3,4$. The dimer density and the trimer density are uniquely
expressed in terms of the monomer density,
\begin{equation}
c_2 = \frac{K_{11}}{F_{11}}\,c_1^2, \qquad
c_3 = \frac{K_{11}}{F_{11}}\,\frac{K_{12}}{F_{12}}\,  c_1^3.
\end{equation}
However, there are two expressions for the $4$-mer density
\begin{eqnarray}
c_4 = \frac{K_{11}}{F_{11}}\,\frac{K_{12}}{F_{12}}\,
\frac{K_{13}}{F_{13}}\, c_1^4   \qquad
c_4 &= \left(\frac{K_{11}}{F_{11}}\right)^2\frac{K_{22}}{F_{22}}\, c_1^4. 
\end{eqnarray}
These two are identical only when the aggregation and the
fragmentation rates satisfy the constraint
\begin{equation}
\label{constraint}
\frac{K_{12}}{F_{12}}\,\frac{K_{13}}{F_{13}} =
\frac{K_{11}}{F_{11}}\,\frac{K_{22}}{F_{22}}.
\end{equation}
Therefore, the detailed balance equations \eqref{balance} have a
solution only for special aggregation and fragmentation rates. The
constraint \eqref{constraint} reflects the fact that there are
multiple paths between two states of the system. For example, a
$4$-mer may be formed by two dimers or by a trimer and a monomer. The
detailed balance condition \eqref{balance} requires that the fluxes
between any two states of the system balance along all possible
paths. This condition leads to constraints of the type
\eqref{constraint}. The rates \eqref{rates} violate the constraint 
\eqref{constraint} as well as infinitely many other constraints.  We
conclude that in general, the detailed balance equations are {\em
overdetermined} --- there are infinitely many constraints like
\eqref{constraint}, and a solution does not necessarily exist.

\section{Large-size asymptotics}
\label{derivation}

We analyze the large-size asymptotic behavior of the polymer
size-density in the thermodynamic phase by introducing the generating
function
\begin{equation}
\label{G:def}
G(z) = \sum_k c_k \,e^{kz}.
\end{equation}
The generating function obeys the differential equation 
\begin{equation}
\label{G:eq}
\frac{(G')^2}{2} - G' = \lambda \,\frac{G-M_0e^z}{1-e^z}   
- \frac{\lambda(G - G')}{2}
\end{equation}
where $'\equiv \frac{d}{dz}$ as follows from \eqref{ck:eq}.  Next, we
shift the generating function by the total density
\begin{equation}
\label{F:def}
G(z) = M_0 + F(z), 
\end{equation}
where $M_0$ is given by \eqref{M0:sol}. With this transformation, the
differential equation \eqref{G:eq} becomes
\begin{equation}
\label{F:eq}
(F')^2-(2+\lambda)F'+1-\lambda+\lambda F\,\frac{e^z+1}{e^z-1} = 0.
\end{equation}
By solving this quadratic equation for $F'$ we find 
\begin{equation}
\label{F:main}
F'(z) = 1+\frac{\lambda}{2} - \sqrt{\Phi(z)}
\end{equation}
with the shorthand notation
\begin{equation}
\label{Phi}
\Phi(z) = \frac{1}{4}\,\lambda^2+2\lambda - \lambda F(z)\,
\frac{e^z+1}{e^z-1}.
\end{equation}

The asymptotic behavior of the size density follows from the singular
behavior of the generating function. For instance, the asymptotic
behavior
\begin{equation}
\label{ck:tail}
c_k \simeq A k^{-\alpha}\,e^{-ak}
\end{equation}
implies that the generating function has the following expansion
\begin{equation}
G(z) = G(a) + G'(a) (z-a) + A\Gamma(1-\alpha)(a-z)^{\alpha-1}+\ldots
\end{equation}
when $z\to a$. Here, it is implicitly assumed that
\hbox{$2<\alpha<3$}. By differentiating this equation and by using $G'
= F'$, we further obtain
\begin{equation}
\label{F:exp}
F'(z) = F'(a) + A\Gamma(2-\alpha)(a-z)^{\alpha-2}.
\end{equation}
We now choose $a$ to be the root of $\Phi(z)$, $\Phi(a) = 0$, and as a
result, equation \eqref{F:main} becomes
\begin{equation}
\label{F:exp2}
F'(z) = 1+\frac{\lambda}{2} - \sqrt{-\Phi'(a)}\, (a-z)^{1/2}+\ldots
\end{equation}
when $z\to a$.  We obtain $F'(a)=1+\lambda/2$ by matching the regular
terms in \eqref{F:exp} and \eqref{F:exp2}, and 
\begin{equation}
\label{aa}
\alpha=5/2, \quad A\Gamma(-1/2) = - \sqrt{-\Phi'(a)}
\end{equation}
by matching the singular terms. Therefore, the asymptotic behavior is 
\eqref{ck:asymp}.

The amplitude $A$ can be expressed in terms of $\Phi'(a)$.
Differentiation of equation \eqref{Phi} yields
\begin{equation}
\Phi'(z) = -\lambda\left[ F'(z)\,
\frac{e^z+1}{e^z-1} - 2F(z)\,\frac{e^z}{(e^z-1)^2}\right].
\end{equation}
We next set $z=a$ and use
\begin{equation}
F'(a)=1+\frac{\lambda}{2}\,, 
\qquad 2F(a)\,\frac{e^a+1}{e^a-1}=4+\frac{\lambda}{2},
\end{equation}
that follows from \eqref{Phi} and $\Phi(a) = 0$ to obtain 
\begin{equation}
\label{Phi:a}
\Phi'(a) = -\frac{\lambda}{2}\left[ (2+\lambda)\,
\frac{e^a+1}{e^a-1} - (8+\lambda)\,\frac{e^a}{e^{2a}-1}\right].
\end{equation}
By using equations \eqref{aa} and \eqref{Phi:a} together with the
identity $\Gamma(-1/2)=-2\Gamma(1/2)=-\sqrt{4\pi}$ we obtain a
relation between the amplitude $A$ and the constant $a$,
\begin{equation}
\label{Aa-gen}
A=\sqrt{\frac{\lambda}{8\pi}\,\frac{e^a+1}{e^a-1} \left[2+\lambda -
(8+\lambda)\,\frac{e^a}{(e^{a}+1)^2}\right]}.
\end{equation}
In particular, $A\simeq\sqrt{3/(16\pi a)}$, when $\lambda\to 1$.

\section{Extremely strong fragmentation}
\label{infinite}

The leading asymptotic behavior in the strong fragmentation limit,
$\lambda\to\infty$, can be obtained analytically.  The steady state
equation \eqref{ck:eq} shows that $c_1=1+O(\lambda^{-1})$ and
$c_2=\lambda^{-1}+O(\lambda^{-2})$, and that in general,
\begin{equation}
\label{hk:sub}
c_k \simeq \left(\frac{2}{\lambda}\right)^{k-1} h_k 
\end{equation}
when $\lambda\to \infty$.

To leading order, this form is consistent with the governing equation
\eqref{ck:eq} when the coefficients $h_k$ satisfy the recursion
equation
\begin{equation}
\label{hk:eq}
(k-1)h_k = \frac{1}{2}\sum_{i+j=k}ijh_ih_j.
\end{equation}
The first two coefficients are $h_1=1$ and $h_2=1/2$. We solve this
recursion using the generating equation \hbox{$H(z)=\sum_k kh_k
e^{kz}$}. Next, we multiply \eqref{hk:eq} by $ke^{kz}$ and sum over
all $k$ to find that the generating function satisfy the nonlinear
differential equation
\begin{equation}
\label{hz-eq}
H'- H = HH'.
\end{equation}
We now integrate this equation and find the implicit solution
$He^{-H}=e^z$. The explicit solution
\begin{equation}
H(z)= \sum_{k=1}^\infty \frac{k^{k-1}}{k!}\,e^{kz}
\end{equation}
follows from the Lagrange inversion formula \cite{wilf}. Therefore,
the coefficients are $h_k = k^{k-2}/k!$ and the leading asymptotic
behavior is \eqref{ck:sol}. The large-size asymptotic behavior 
\begin{equation}
\label{ck:asymp-detail}
c_k = \frac{e}{\sqrt{2\pi}}\,k^{-5/2}\left(\frac{2e}{\lambda}\right)^{k-1}, 
\end{equation}
when $k\gg 1$, obtained using the Stirling formula \hbox{$n!\sim
\sqrt{2\pi n}n^ne^{-n}$}, is consistent with the generic asymptotic
behavior \eqref{ck:asymp}.


\begin{thebibliography}{99}
\bibitem{note} The rate equations \eqref{master} are mean-field and
    therefore they could be invalid for low-dimensional
    systems. Nevertheless, these master equations can be justified in
    a number of experimentally relevant conditions including for
    example three dimensional systems, dilute conditions, and
    well-mixed solutions.
\end{thebibliography}
\end{document}